# An Improved Fixed Switching Frequency Direct Torque Control of Induction Motor Drives Fed by Direct Matrix Converter

Nabil Taïb and Toufik Rekioua
LTII Laboratory, Electrical Engineering Department
University of A. Mira, Targua Ouzemour,
Bejaia, 06000, Algeria.
taib_nabil@yahoo.fr, to_reki@yahoo.fr

Bruno François
L2EP Laboratory
Central School of Lille, BP48,
Villeneuve d Ascq, Lille 59651, France
bruno.francois@ec-lille.fr

*Abstract—* a few papers have been interested by the fixed switching frequency direct torque control fed by direct matrix converters, where we can find just the use of direct torque controlled space vector modulated method. In this present paper, we present an improved method used for a fixed switching frequency direct torque control (DTC) using a direct matrix converter (DMC). This method is characterized by a simple structure, a fixed switching frequency which causes minimal torque ripple and a unity input power factor. Using this strategy, we combine the direct matrix converters advantages with those of direct torque control (DTC) schemes. The used technique for constant frequency is combined with the input current space vector to create the switching table of direct matrix converter (DMC). Simulation results clearly demonstrate a better dynamic and steady state performances of the proposed method.

## I. INTRODUCTION

The direct ac-ac matrix converter, appeared in 1980 |1]-[3], has recently received a considerable attention these last years because of its numerous merits such as no dc link capacitor with limited life time, the bi-directional power flow control, the sinusoidal input output waveforms and adjustable input power factor.

Moreover, the direct matrix converter (DMC) is compact sight because of the lack of the dc-link bus capacity used for the energy storage. Comprehensive researches on direct matrix converter, so far, have been focused on modulation methods [4]-[7] and some researches on direct matrix converter topologies [8]-[9].

After, with the improvement of the hardware microcontrollers and bidirectional switches, the focused of the research will be on the applications of these kinds of converter such as variable speed drives and some other applications.

The Direct Torque Control (DTC) methods of alternating machines appeared in the second half of 1980 like a competitive of the traditional methods, based on a PWM supplies and a decoupling of flux and torque by the magnetic field orientation control (FOC).

The direct torque control scheme for direct matrix converter was initially presented in [10]. The generation of the voltage vectors required to implement the DTC of induction motors under unity input power factor constraint was allowed. However, the DTC scheme using a switching table has some fatal drawbacks. Switching frequency varies according to the motor speed and the hysteresis bands of the torque and flux, a large torque ripple is generated in a low speed range because of the small back electromotive force of the induction motor, and high control sampling time is required to achieve good performances.

Although, several methods to solve these problems have been presented [11]-[13], these methods are designed for a conventional inverter drive system. On the contrary, for the matrix converter a few solutions have been presented [14], which uses the direct torque controlled space vector modulated method (DTC-SVM).

In this paper, a new fixed switching frequency DTC scheme using one switching table for DMC is presented, which enables to minimize torque ripple under a unity input power factor constraint. The proposed control scheme of the fixed switching frequency DTC of an induction motor drives fed by DMC shown in Fig. 1, which consists of the basic DTC switching table, detection of input voltage angle and a high frequency triangular waveform. The effectiveness of the proposed control scheme is demonstrated through simulation results.

## II. DESCRIPTION OF THE GLOBAL SYSTEM

The classic DTC method is based on this algorithm:

- At each period, the lines currents and voltages are measured;
- The stator vector flux components are constructed;
- The torque estimation is, now, possible to be performed.

The flux and torque reference values are compared with their actual values, and the resulting values are fed in two-level and three-level hysteresis comparators respectively. The outputs of both stator flux and torque comparators respectively ($\phi$, $\tau$), together with the position of the stator flux are used as inputs of the basic DTC switching look up table (Table I). The position of the stator flux is divided into six different sections ($\theta$(i)) as shown in fig. 1.

TABLE I. BASIC DTC SWITCHING TABLE

| $\phi$ | $\tau$ | $\theta$(1) | $\theta$(2) | $\theta$(3) | $\theta$(4) | $\theta$(5) | $\theta$(6) |
|---|---|---|---|---|---|---|---|
| 1 | 1 | V2 | V3 | V4 | V5 | V6 | V1 |
| 1 | 0 | V7 | V0 | V7 | V0 | V7 | V0 |
| 1 | -1 | V6 | V1 | V2 | V3 | V4 | V5 |
| 0 | 1 | V3 | V4 | V5 | V6 | V1 | V2 |
| 0 | 0 | V0 | V7 | V0 | V7 | V0 | V7 |
| 0 | -1 | V5 | V6 | V1 | V2 | V3 | V4 |

Fig. 2 shows the global proposed drive scheme using the constant frequency DTC, based on the classic table and the method described in the section IV.

## III. DIRECT MATRIX CONVERTER THEORY

The three phase direct matrix converter is the nine bidirectional switches converter which allows connecting any output phase to any input phase as shown in Fig. 2.

Normally, the direct matrix converter is fed by voltage source; for this reason, the input terminal should not be shorted. On the other hand, the load has generally an inductive nature, thus, the output phase must never be opened. The output phase voltages $v_o(t)$ and the input phase currents $i_i(t)$ are defined by (1) and (2):

Figure 3. The proposed fixed frequency DTC scheme using matrix converter

$$v_o(t) = M(t) \cdot v_i(t) \qquad (1)$$

$$i_i(t) = M^T(t) \cdot i_o(t) \qquad (2)$$

Where:

$$v_i = \begin{bmatrix} V_a \\ V_b \\ V_c \end{bmatrix} \; ; v_o = \begin{bmatrix} V_A \\ V_B \\ V_C \end{bmatrix} \; ; i_i = \begin{bmatrix} i_a \\ i_b \\ i_c \end{bmatrix} \; ; i_o = \begin{bmatrix} i_A \\ i_B \\ i_C \end{bmatrix}$$

With the duty cycles matrix $M(t)$ is defined by (3):

$$M(t) = \begin{bmatrix} m_{aA}(t) & m_{bA}(t) & m_{cA}(t) \\ m_{aB}(t) & m_{bB}(t) & m_{cB}(t) \\ m_{aC}(t) & m_{bC}(t) & m_{cC}(t) \end{bmatrix} \qquad (3)$$

Where $m_{ij}(t)$ is the conduction duration of the bidirectional switching $S_{ij}$ and $M^T(t)$ is the transpose of the duty cycles matrix $M(t)$.

For 3-phase/3-phase matrix converters, there are 27 possible switching vectors which are summarized in Table II.

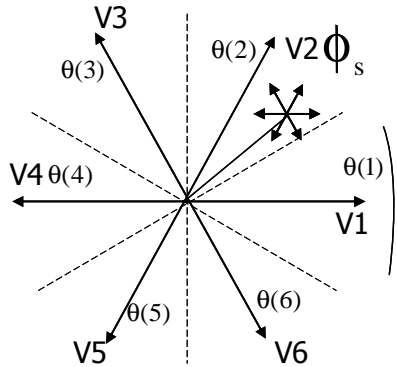

Figure 1. VSI output line to neutral voltage vector and corresponding stator flux variation

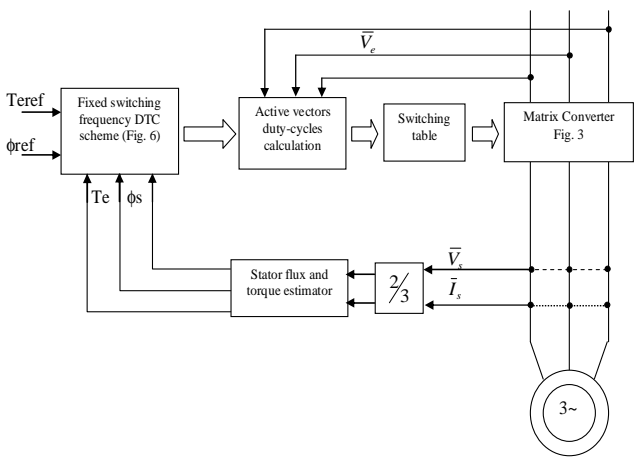

Figure 2. The proposed fixed frequency DTC scheme using matrix converter

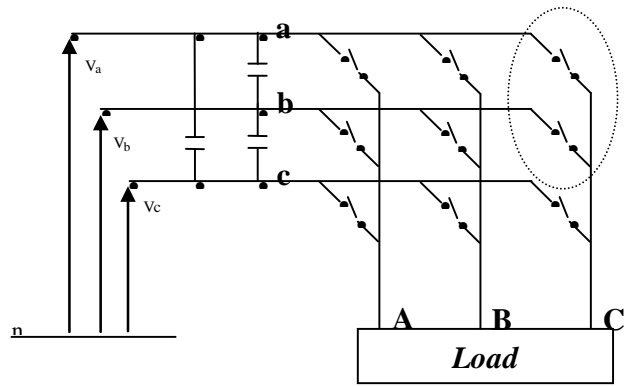

Figure 3. Conventional Direct matrix converter

For the combinations from 1 to 3, all output phases are connected to the same input phase. In the next 18 combinations, two output phases are connected to the same input phase. In the last 6 combinations, each output phase is connected to a different input phase.

TABLE II.　SWITCHING STATES AND CORRESPONDING LINE-TO-LINE OUTPUT VOLTAGES AND INPUT PHASE CURRENTS OF MATRIX CONVERTER.

| No | A | B | C | $S_{aA}$ | $S_{bA}$ | $S_{cA}$ | $S_{aB}$ | $S_{bB}$ | $S_{cB}$ | $S_{aC}$ | $S_{bC}$ | $S_{cC}$ | $U_{AB}$ | $U_{BC}$ | $U_{CA}$ | $i_a$ | $i_b$ | $i_c$ |
|---|---|---|---|---|---|---|---|---|---|---|---|---|---|---|---|---|---|---|
| 1 | a | a | a | 1 | 0 | 0 | 1 | 0 | 0 | 1 | 0 | 0 | 0 | 0 | 0 | 0 | 0 | 0 |
| 2 | b | b | b | 0 | 1 | 0 | 0 | 1 | 0 | 0 | 1 | 0 | 0 | 0 | 0 | 0 | 0 | 0 |
| 3 | c | c | c | 0 | 0 | 1 | 0 | 0 | 1 | 0 | 0 | 1 | 0 | 0 | 0 | 0 | 0 | 0 |
| 4 | a | c | c | 1 | 0 | 0 | 0 | 0 | 1 | 0 | 0 | 1 | -Uca | 0 | Uca | $i_A$ | 0 | -$i_A$ |
| 5 | b | c | c | 0 | 1 | 0 | 0 | 0 | 1 | 0 | 0 | 1 | Ubc | 0 | -Ubc | 0 | $i_A$ | -$i_A$ |
| 6 | b | a | a | 0 | 1 | 0 | 1 | 0 | 0 | 1 | 0 | 0 | -Uab | 0 | Uab | -$i_A$ | $i_A$ | 0 |
| 7 | c | a | a | 0 | 0 | 1 | 1 | 0 | 0 | 1 | 0 | 0 | Uca | 0 | -Uca | -$i_A$ | 0 | $i_A$ |
| 8 | c | b | b | 0 | 0 | 1 | 0 | 1 | 0 | 0 | 1 | 0 | -Ubc | 0 | Ubc | 0 | -$i_A$ | $i_A$ |
| 9 | a | b | b | 1 | 0 | 0 | 0 | 1 | 0 | 0 | 1 | 0 | Uab | 0 | -Uab | $i_A$ | -$i_A$ | 0 |
| 10 | c | a | c | 0 | 0 | 1 | 1 | 0 | 0 | 0 | 0 | 1 | Uca | -Uca | 0 | $i_B$ | 0 | -$i_B$ |
| 11 | c | b | c | 0 | 0 | 1 | 0 | 1 | 0 | 0 | 0 | 1 | -Ubc | Ubc | 0 | 0 | $i_B$ | -$i_B$ |
| 12 | a | b | a | 1 | 0 | 0 | 0 | 1 | 0 | 1 | 0 | 0 | Uab | -Uab | 0 | -$i_B$ | $i_B$ | 0 |
| 13 | a | c | a | 1 | 0 | 0 | 0 | 0 | 1 | 1 | 0 | 0 | -Uca | Uca | 0 | -$i_B$ | 0 | $i_B$ |
| 14 | b | c | b | 0 | 1 | 0 | 0 | 0 | 1 | 0 | 1 | 0 | Ubc | -Ubc | 0 | 0 | -$i_B$ | $i_B$ |
| 15 | b | a | b | 0 | 1 | 0 | 1 | 0 | 0 | 0 | 1 | 0 | -Uab | Uab | 0 | $i_B$ | -$i_B$ | 0 |
| 16 | c | c | a | 0 | 0 | 1 | 0 | 0 | 1 | 1 | 0 | 0 | 0 | Uca | -Uca | $i_C$ | 0 | -$i_C$ |
| 17 | c | c | b | 0 | 0 | 1 | 0 | 0 | 1 | 0 | 1 | 0 | 0 | -Ubc | Ubc | 0 | $i_C$ | -$i_C$ |
| 18 | a | a | b | 1 | 0 | 0 | 1 | 0 | 0 | 0 | 1 | 0 | 0 | Uab | -Uab | -$i_C$ | $i_C$ | 0 |
| 19 | a | a | c | 1 | 0 | 0 | 1 | 0 | 0 | 0 | 0 | 1 | 0 | -Uca | Uca | -$i_C$ | 0 | $i_C$ |
| 20 | b | b | c | 0 | 1 | 0 | 0 | 1 | 0 | 0 | 0 | 1 | 0 | Ubc | -Ubc | 0 | -$i_C$ | $i_C$ |
| 21 | b | b | a | 0 | 1 | 0 | 0 | 1 | 0 | 1 | 0 | 0 | 0 | -Uab | Uab | $i_C$ | -$i_C$ | 0 |
| 22 | a | b | c | 1 | 0 | 0 | 0 | 1 | 0 | 0 | 0 | 1 | Uab | Ubc | Uca | $i_A$ | $i_B$ | $i_C$ |
| 23 | a | c | b | 1 | 0 | 0 | 0 | 0 | 1 | 0 | 1 | 0 | -Uca | -Ubc | -Uab | $i_A$ | $i_C$ | $i_B$ |
| 24 | b | a | c | 0 | 1 | 0 | 1 | 0 | 0 | 0 | 0 | 1 | -Uab | -Uca | -Ubc | $i_B$ | $i_A$ | $i_C$ |
| 25 | b | c | a | 0 | 1 | 0 | 0 | 0 | 1 | 1 | 0 | 0 | Ubc | Uca | Uab | $i_B$ | $i_C$ | $i_A$ |
| 26 | c | a | b | 0 | 0 | 1 | 1 | 0 | 0 | 0 | 1 | 0 | Uca | Uab | Ubc | $i_C$ | $i_A$ | $i_B$ |
| 27 | c | b | a | 0 | 0 | 1 | 0 | 1 | 0 | 1 | 0 | 0 | -Ubc | -Uab | -Uca | $i_C$ | $i_B$ | $i_A$ |

The first 21 states can be used with DTC scheme. There are 18 configurations named "active vectors" allow the connection of two output phases to the same input phase. In the 3 remainder configurations named "zero vectors" all output phases are connected to the same input phase [10]-[11].

## IV. FIXED SWITCHING FREQUENCY DTC PRINCIPLES USING DIRECT MATRIX CONVERTER

### A. Fixed Switching Frequency DTC

The stator flux and the electromagnetic torque expressions are given by (1) and (2) respectively:

$$\phi_s = \int_0^t V_s dt \quad (1)$$

$$T_e = \frac{3}{2} p \frac{L_m}{\sigma L_s L_r} |\phi_s||\phi'_r| \sin\gamma \quad (2)$$

Where,

| | |
|---|---|
| $V_s$ | Stator supply voltage vector |
| p | Number of pole pairs |
| $L_m$ | Mutual inductance |
| $L_s, L_r$ | Stator and rotor self inductance |
| $\sigma$ | Total leakage factor |
| $\phi_s$ | Stator flux linkage |
| $\phi'_r$ | Rotor flux linkage expressed in the stationary frame |
| $\gamma$ | Angle between stator and rotor flux linkages |

The electromagnetic torque is a sinusoidal function of $\gamma$, the angle between the stator $\phi_s$ and $\phi'_r$ space vector. The magnitude of the stator flux is normally kept constant and the motor torque controlled by means of the angle $\gamma$. The rotor time constant of the standard induction machine is typically larger than 100ms; thus the rotor flux is stable and variations in the rotor flux are slow compared with stator flux. It is therefore possible to achieve the required torque very effectively by rotating the stator flux vector, means the stator voltage vector, directly in a given direction as fast as possible.

The variable switching frequency in the basic DTC is due to the variation of the time taken for the torque error to achieve the upper and lower hysteresis bands. The waveform and the dynamic of the torque error slopes are highly dependant on operating conditions. Consequently, the torque ripple will remain high, even with small hysteresis band (Fig. 4).

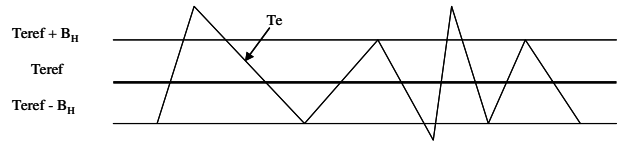

Figure 4.　Torque error causing variable switching frequency

By assuming a fixed gradient of the electromagnetic torque 'Te' over each sampling period, the suitable form of the reference should be triangular. So, we will superpose the torque error with high frequency triangular waveform whose parameters as the magnitude $A_{tr}$ and the frequency $f_{tr}$, depends on the torque dynamic and band width of the hysteresis torque $B_H$ (Fig. 5).

In this case, the esteemed torque variation during a half period of the triangular signal should not exceed the difference between the maximum of the upper limit and the minimum of the lower one like it is shown in Fig. 6.

From Fig. 5, and assuming that the triangular waveform parameters must not exceed the maximum of the electromagnetic torque variations, it can be demonstrated that:

$$\left\|\frac{dT_e}{dt}\right\|_{max} \leq \frac{4(A_{tr} + B_H)}{T_{tr}} \quad (3)$$

Where $T_{tr}$ is the triangular waveform period.

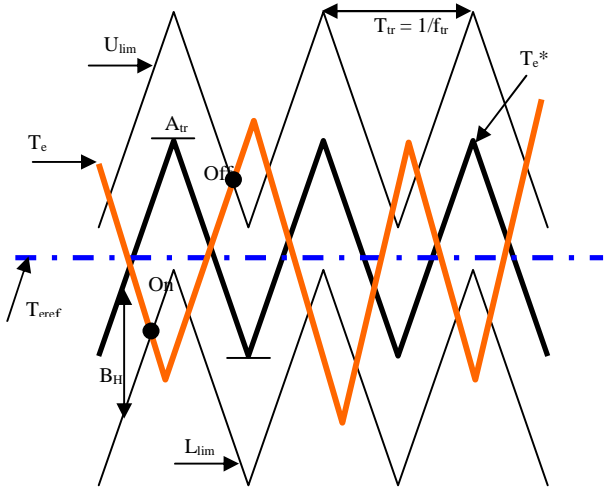

Figure 5.  Switches states determination with modulated torque reference

It is well known that we can express the electromagnetic torque in Park frame with stator and rotor currents only (4):

$$T_e = \frac{3}{2} L_m \left( i_{ds} i_{qr} - i_{qs} i_{dr} \right) \quad (4)$$

Since the frequency of the rotor current is very smaller than the stator one, so, it can be concluded that the dynamic of the torque is directly affected by the dynamic of the current and vice-versa.

*B. Fixed Switching Frequency DTC using Matrix Converter*

From the basic DTC, it appears that the conventional switching table (Table I) is not able to generate the full vector of nine signal gates, which allows the direct matrix converter control, because we have nine bidirectional switches to control in the DMC and each output phase can be connected to any input phase. Then, the most modulation technique used for the DMC is the indirect space vector modulation (ISVM) method [5]-[6]. In this technique, we consider the DMC as a two stages power converter, the first stage is a rectifier current source, the second stage is a voltage source inverter. So, the vector generated from the Table I is assimilated to the SVM applied to the inverter stage of direct matrix converter. To have a fully control of the DMC, we need to use the input current modulator for the rectification stage like it is used in the (ISVM) method. Like this, , we will be able to generate the switching table for direct matrix converter.

In principle, the proposed control technique selects, at each sampling period, the proper switching configuration which allows the compensation of instantaneous errors in flux and torque, under the constraint of unity input power factor. This last requirement of the direct matrix converter input side is intrinsically satisfied if we assume that the reference space vector of the input current is the input voltage vector. When an active vector is selected from the DTC switching table, the input current reference vector is lying in one of the six sectors like it is shown in Fig. 7. The duty cycles used for the commutation sequences, which will be defined later, are computed as in (5):

$$d_\gamma = \sin\left(\frac{\pi}{3} - \theta_{in}\right)$$
$$d_\delta = \sin(\theta_{in}) \quad (5)$$

Where $\theta_{in}$ is the angle in the switching hexagon of the input current reference vector with the right axis of the sector where lies the reference current space vector.

Since the zero vectors are used in DTC to maintain the torque constant, the applied switching sequences are not including the zero sequence (000 or 111) during the switching period. Then, the zero switching vector is applied during the duty cycle $d_0$ computed as in (6):

$$d_0 = 1 - \left(d_\delta + d_\gamma\right) \quad (6)$$

So, the duty cycles may be adjusted to have two vectors to apply during the whole switching period. To solve this problem, an approach is used to adjust the duty cycles. If we assume that the new duty cycles are $d'_\gamma$ and $d'_\delta$, we can resize the duty cycles $d_\gamma$ and $d_\delta$ using the expressions given by (7):

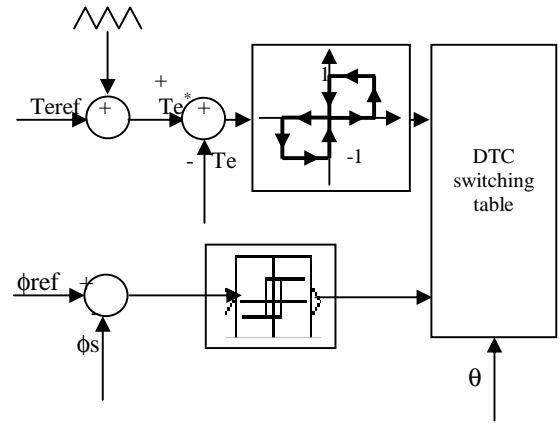

Figure 6.  Proposed fixed frequency DTC scheme.

$$d'_\gamma = \frac{d_\gamma}{d_\gamma + d_\delta}$$
$$d'_\delta = \frac{d_\delta}{d_\gamma + d_\delta} \quad (7)$$

To explain the implementation of the direct matrix converter switching table, we can refer to an example. If we consider that $V_1$ ($V_1$ = [100]) is the output voltage vector selected by the DTC basic algorithm in a given switching period, according to the sector in which the input current reference vector is lying, we can generate the switching sequences. For example, if this sector is "I", the two switching sequences able to apply are "*abb*" and "*acc*" (see Fig. 4), i.e. for the first sequence, the first, the second and the third output phases are connected respectively to the first, second and second input phases, during the duty-cycle $d'_\gamma$.

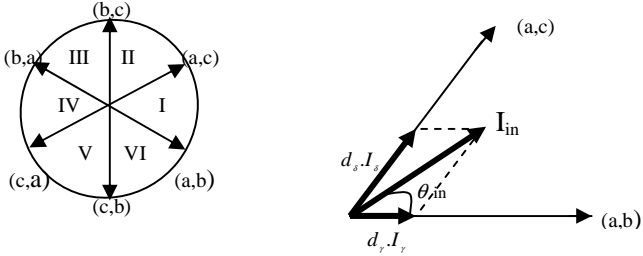

Figure 7.  Input current reference vector hexagon

Finally, the duration of each sequence is calculated by multiplying the corresponding duty-cycle $T_\gamma$ and $T_\delta$ to the switching period as in (7):

$$T_\gamma = d'_\gamma \cdot T_c$$
$$T_\delta = d'_\delta \cdot T_c \quad (7)$$

Where, $T_c$ is the switching period.

So, following the above reasoning, the proposed switching table (Table III) will be created.

TABLE III.    PROPOSED DMC SWITCHING TABLE FOR FIXED SWITCHING FREQUENCY DTC

| $\theta_{in}$ / $V_i$(dtc) | I | II | III | IV | V | VI |
|---|---|---|---|---|---|---|
| V1=[100] | abb, acc | acc, bcc | bcc, baa | baa, caa | caa, cbb | cbb, abb |
| V2=[110] | aab, aac | aac, bbc | bbc, bba | bba, cca | cca, ccb | ccb, aab |
| V3=[010] | bab, cac | cac, cbc | cbc, aba | aba, aca | aca, bcb | bcb, bab |
| V4=[011] | baa, caa | caa, cbb | cbb, abb | abb, acc | acc, bcc | bcc, baa |
| V5=[001] | bba, cca | cca, ccb | ccb, aab | aab, aac | aac, bbc | bbc, bba |
| V6=[101] | aba, aca | aca, bcb | bcb, bab | bab, cac | cac, cbc | cbc, aba |
| V0=[000] | bbb, ccc | ccc, ccc | ccc, aaa | aaa, aaa | aaa, bbb | bbb, bbb |
| V7=[111] | aaa, aaa | aaa, bbb | bbb, bbb | bbb, ccc | ccc, ccc | ccc, aaa |

## V. SIMULATION RESULTS AND DISCUSSIONS

The proposed drive system was tested by digital simulations, under MATLAB/SIMULINK package program using the scheme of Fig. 2, with an aim of checking and comparing its performances with the same carried out switching table (Table III) without modulate the reference torque.

The test machine is a 1.5 kW, 2-pole, 380/220V and 50Hz cage induction having the following parameters:

Rs=4.85 Ω; Rr=3.805Ω; Ls=Lr=.274 H; Lm=.258 H

The simulation has been carried out assuming that the sampling period is 50μs and ideal switching devices.

### A. Steady State Behavior

The obtained results for a 10 Nm torque reference and 1.14 Wb flux reference, in terms of electromagnetic torque and stator flux, are presented in Fig. 8 and 9.

It can be seen from Fig. 9a, that magnitudes of stator flux and electromagnetic torque follow their references. Fig. 9b shows that the torque ripple are reduced by using the proposed control method, compared to the case where the switching frequency is not constant (see Fig. 8b), which is one of the main reasons for which the fixed switching frequency DTC was designed.

Thus, the good performances of the drive system with regard to the basic DTC technique implementation are achieved. Moreover, the stator current presents good wave forms as it is shown in Fig. 10, 11 and 12 and also the stator flux trajectory which is circular (Fig. 13)

We can be seen clearly, from Fig. 10 and 12, that the quality of the stator current waveform is better in case of the constant frequency DTC than that without constant frequency.

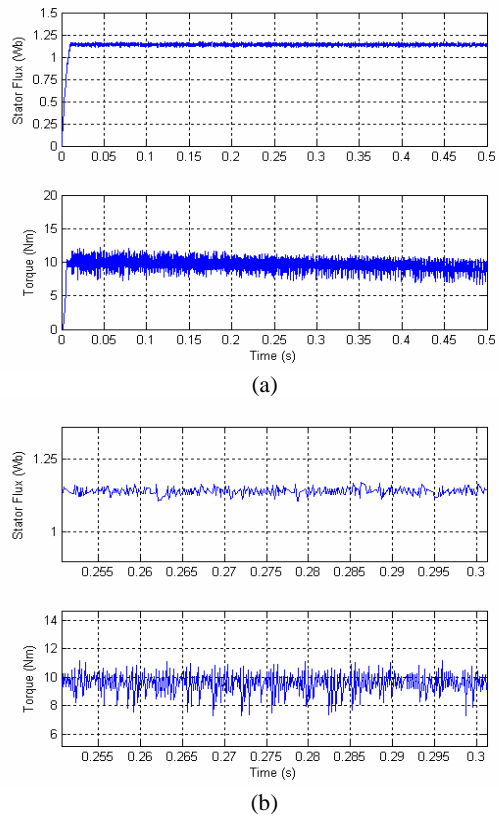

Figure 8.  Flux and torque response without fixed switching frequency DTC

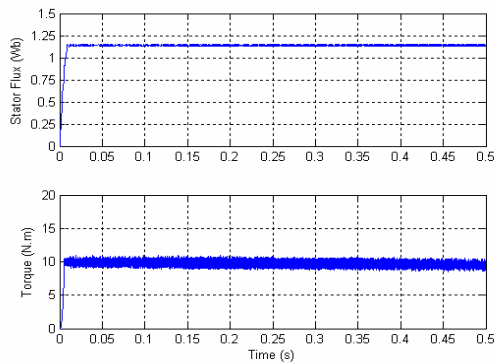

(a)

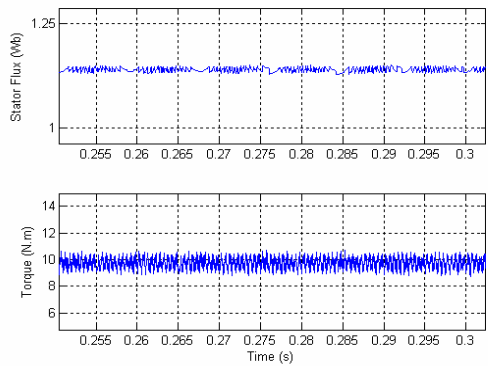

(b)

Figure 9. Flux and torque response with the proposed DTC scheme

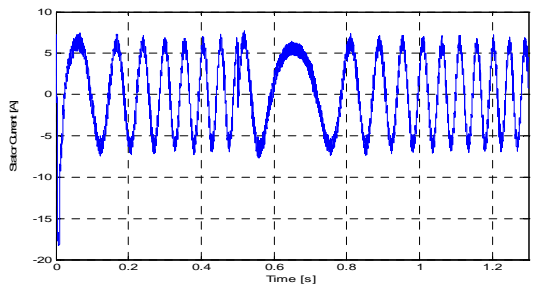

(a)

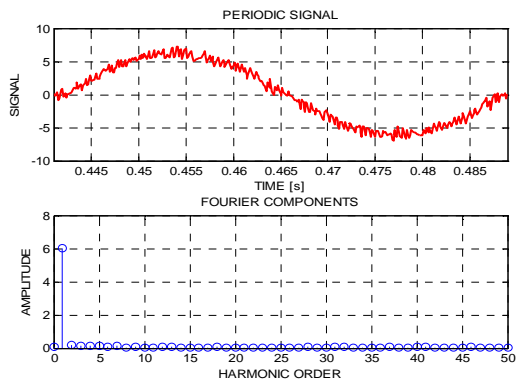

(b)

Figure 10. Stator current and its harmonic spectrum without constant frequency

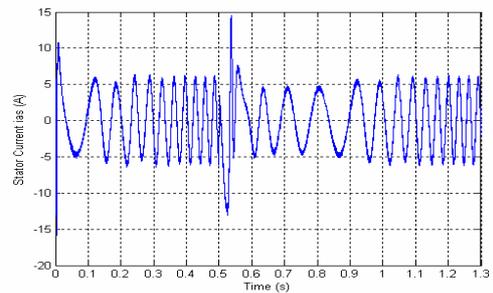

(a)

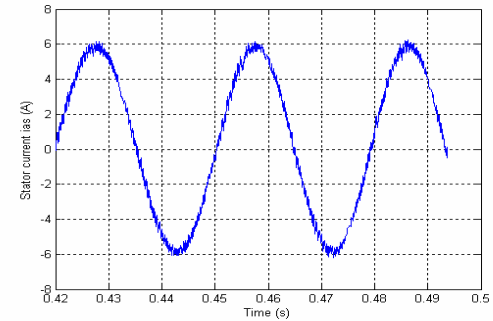

(b)

Figure 11. Stator current with the proposed DTC scheme

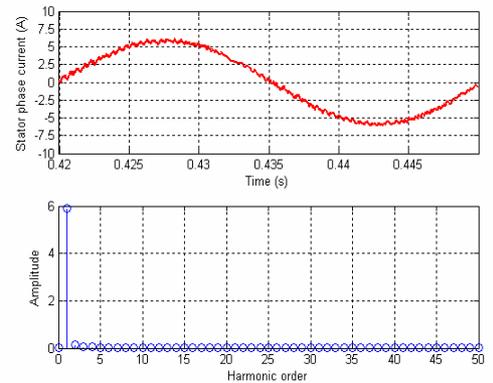

Figure 12. Stator current harmonic spectrum with proposed method

The form of the input current Fig. 17 is also good. The validity of the proposed strategy is confirmed as the operation of the converter is at unity input power-factor. It offers also the advantage of carrying out a minimal number of switchings as well as a cell switching frequency, about 5 kHz, lower than with the conventional DTC, which can be higher than 20 kHz.

*B. Dynamic Behavior*

The dynamic performances of the proposed control method are tested for a step torque command from +10 Nm to -10 Nm and shown in Fig. 14.

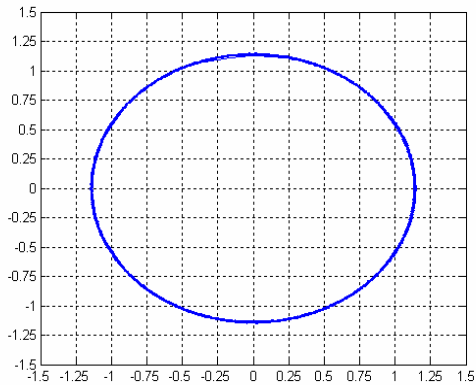

Figure 13. Stator flux in a polar coordinate

From Fig.15 and 16, we can see the improvement guarantee by the proposed method in term of maintaining constant the DTC switching frequency and then reducing the torque ripple. In case of not constant frequency, the electromagnetic torque dynamic is variable as shown in Fig. 15. These variations are from 11 % to 34 % for the torque slopes with a frequency varies between 3.334 kHz and 10 kHz for the variable frequency DTC scheme. For the proposed DTC scheme, it can be seen that the torque dynamic is constant equals to 10% with constant frequency equals to 5 kHz (Fig. 16). Exploring the Fig. 11 and 12, we can confirm the relation between the dynamic of stator currents and that of electromagnetic torque, where the ripples are reduced with the proposed constant frequency DTC.

The change in the input current direction in case of an inverse sense of the machine speed at 0.50s is shown on Fig. 15 and 16 where the dynamic of the electromagnetic torque follow the reference modulated with a constant gradient and also causes a reduced torque ripple. Thus, in these operating conditions the direct matrix converter transits the generated electric power by the induction machine to the grid with closely sinusoidal waveforms and unity input power factor (Fig. 17). This demonstrates the bidirectional power flow of the direct matrix converter.

## VI. CONCLUSION

A new constant switching frequency DTC of induction motor drives, in which a direct matrix converter is used for supplying the system, is proposed. A new switching table, based on the requirement of the induction machine DTC and those of the converter, has been defined. The proposed control scheme has been tested as well in the steady-state as in the dynamic transients. The torque, the flux and current waveforms emphasize the effectiveness of the control scheme. The proposed technique using the fixed switching frequency DTC helped minimize torque ripple. The results show that, during the regenerative breaking, the drive system acts as a nearly sinusoidal unity input power factor generator.

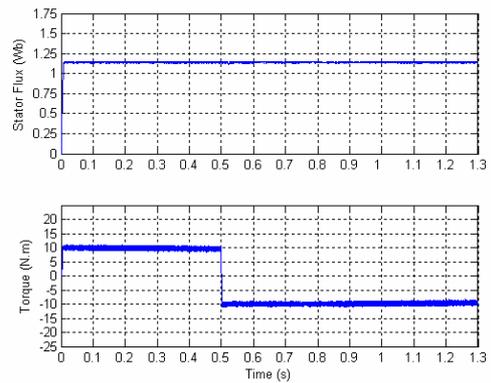
(a)

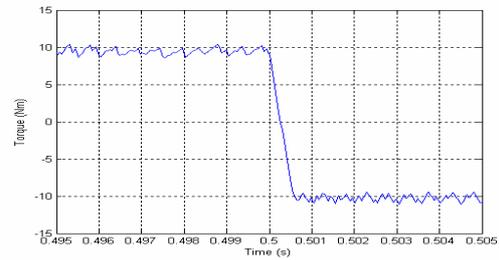
(b)

Figure 14. Electromagnetic torque in dynmic state with proposed DTC

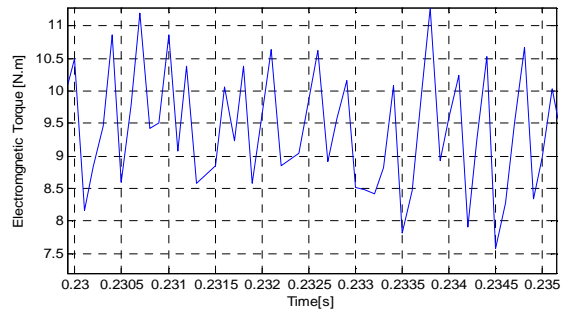

Figure 15. Electromagnetic torque dynamic without constant frequency DTC

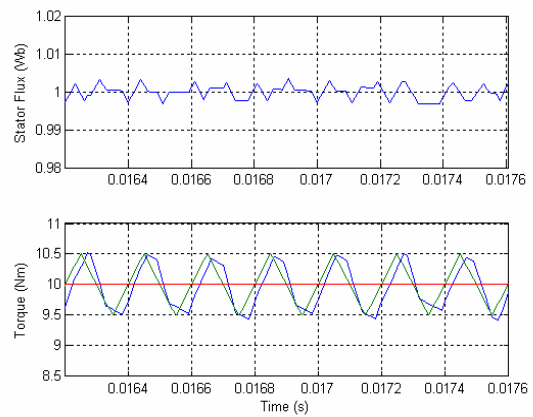

Figure 16. Stator flux and electromagnetic torque variations with fixed frequency DTC

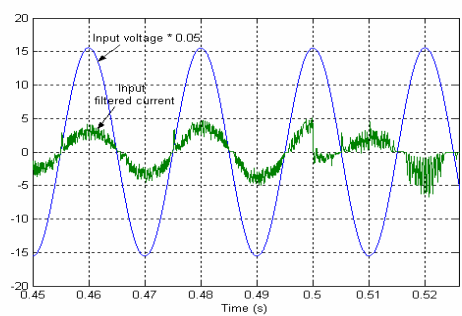

(a)

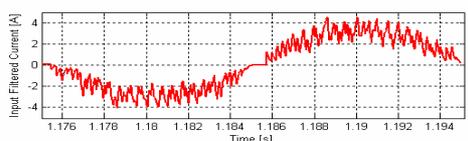

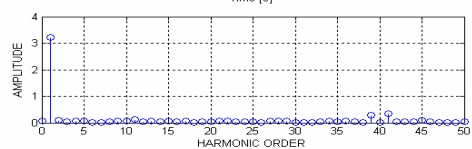

(b)

Figure 17. Input phase voltage and current of direct matrix converter

## AUTHORS PROFILE

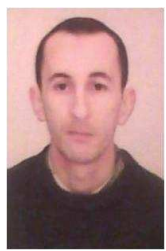

Nabil Taib (1977), received his Engineering Degree and Master Degree on Electrical Control Systems in 2001 and 2004 respectively from the University of Bejaia (Algeria). Since 2005 he is preaparing his Doctoral thesis at the same University where the focused research is on the matrix converters and their applications. From 2009, he is assistant master in the Electrical Engineering Department at the University A. Mira of Bejaia (Algeria). He is interrested now, by the applications of the matrix converters on the renewable energy systems.

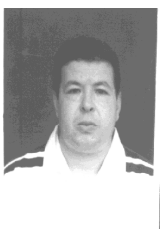

Toufik Rekioua (1962), received his Egineering Degree from the Polytechnical National School (Algeria) and earned the Doctoral Degree from I.N.P.L of Nancy (France) in 1991. Since 1992, he is assistant professor at the Electrical Engineering department-University A. Mira of Bejaia (Algeria). His research activities have been devoted to several topics: control of electrical drives, modeling and control of A.C machines, the renewable energy systems.

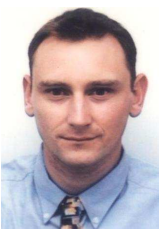

Bruno Francois, received the Ph.D. and "Habilitation à Diriger Recherches" degrees from the University of Lille, Lille, France, in 1996 and 2003, respectively. He is an Assistant Professor with the Department of Electrical Engineering, Ecole Centrale de Lille. He is a member of Laboratory of Electrical Engineering (L2EP), Lille. He is currently working on the design of modulation and control systems for multilevel converters and also the development of next-generation power systems.